\documentclass[reprint,
superscriptaddress,
 amsmath, amssymb, aps, pra,
floatfix,
]{revtex4-2}
\usepackage[margin=0.83in]{geometry}
\usepackage{graphicx}
\usepackage{amsmath}
\usepackage{comment,physics}
\usepackage{calrsfs}
\usepackage{simpler-wick}
\DeclareMathAlphabet{\pazocal}{OMS}{zplm}{m}{n}
\usepackage{resizegather}

\begin{document}
\title{Matter-Field Exchange Generates Entanglement, Not Classical Gravity}
\author{Nicetu Tibau Vidal}
\email{nicetutv@gmail.com}
\affiliation{Clarendon Laboratory, Department of Physics, University of Oxford, Oxford OX1 3PU, United Kingdom}
\affiliation{QICI Quantum Information and Computation Initiative, Department of Computer Science, The University of Hong Kong, Pok Fu Lam Road, Hong Kong}

\author{Aditya Varna Iyer}
\affiliation{Clarendon Laboratory, Department of Physics, University of Oxford, Oxford OX1 3PU, United Kingdom}

\begin{abstract}
Aziz and Howl have argued that a hybrid theory with quantum matter and a classical gravitational field can generate entanglement between two massive systems.  In their construction, the classical metric does not itself enter a quantum superposition; instead, the branch-dependent contribution appears at fourth order through propagators of the quantum matter field in a fixed classical gravitational potential.  We analyse this mechanism within the same perturbative QFT framework and show that the effect should not be interpreted as classical gravity mediating entanglement in the sense relevant to BMV-type witnesses.  The non-separable term relies on a quantum-matter exchange channel between the two interferometers and is present only when the two systems are modelled as excitations of the same matter field.  If distinct, non-interconverting matter fields describe the systems, the corresponding cross-propagator is absent and the Aziz--Howl entangling diagram vanishes.  The effect is instead a matter-sector exchange channel: it relies on coherent propagation amplitudes of the quantum matter field between the two interferometers, together with postselection onto the original localised branch subspace. Moreover, the inference of entanglement is made after projecting the full QFT evolution onto a restricted final subspace containing the original localised $N$-particle branch states.  This projection removes precisely the sectors that would record matter-field contamination, mode deformation, or exchange between the two interferometers.  We therefore argue that the Aziz--Howl mechanism lies outside the assumptions of the BMV witness: it is a matter-sector cross-talk effect in a classical background, not entanglement mediated by classical gravitational degrees of freedom. Having identified the channel responsible for the Aziz–Howl contribution, we predict that it can be eliminated by using distinct, non-interconverting matter species in the two interferometers, or by inserting a barrier that suppresses matter-field propagation between them. Either modification removes the matter-sector exchange channel and should therefore prevent the Aziz–Howl mechanism from generating entanglement.
\end{abstract}

\maketitle

\section*{Introduction}

The recent proposal by Aziz and Howl (AH)~\cite{azizClassicalTheoriesGravity2025} argues that a hybrid model, in which matter is treated quantum mechanically while gravity remains classical, can nevertheless generate entanglement between two massive systems.  Their central claim is not that a branch-dependent classical gravitational potential directly entangles the systems, but rather that once matter is described quantum-field-theoretically, the classical gravitational interaction can open quantum communication channels via matter-field propagators.  In their calculation, the first non-separable contribution appears at fourth order in perturbation theory and, after a long-time delta-function approximation, produces a branch-dependent $1/r$ kernel analogous to the Newtonian potential.  If interpreted as entanglement mediated by classical gravity, this claim would weaken the Bose--Marletto--Vedral (BMV) witness~\cite{boseSpinEntanglementWitness2017,marlettoGravitationallyInducedEntanglement2017}, according to which observation of gravity-mediated entanglement implies non-commuting, and hence quantum, gravitational degrees of freedom.%

In this article, we review the AH mechanism within their own perturbative QFT framework.  We do not dispute the general field-theoretic point that quantum matter fields in a classical background possess propagators, nor that such propagators can appear in diagrams generated by a classical gravitational coupling.  The crucial question is instead what physical channel the fourth-order AH term represents.  We argue that the non-separable contribution is not classical gravity transmitting quantum information.  Rather, it is a quantum matter-sector channel enabled by the common matter field used to describe the two interferometers.  In the long-time approximation used to obtain the unsuppressed $1/r$ behaviour, the matter propagator is effectively projected onto its mass shell.  The resulting contribution is therefore more naturally interpreted as resonant matter-field propagation between the two localised modes---a leakage or contamination channel---than as a purely virtual exchange mediated by classical gravitational degrees of freedom. By noting that the production of entanglement relies on the physical exchange of identical particles via diffusion between the two interferometers and on postselection of the final states, we show that the interaction violates the assumptions underlying all BMV arguments and should be explicitly prevented in any experimental setup by incorporating barrier-like structures to block such leakage. Moreover, due to the nature of the entangling effect, we comment on the possible breakdown of this mechanism when collective spin states are considered.  

Despite AH having several replies in the literature \cite{gundhiCanClassicalTheories2026,diosiNoClassicalGravity2026,schneiderDemonstrationThatClassical2026,xueAzizHowlsGravityInduced2026,tangMatterMediatedEntanglementClassical2025,biagioSimpleReasonWhy2026,marlettoClassicalGravityCannot2025}, we specifically work within the framework of AH, addressing its mechanism and setup directly, without relying on arguments and results from other frameworks. We complement and expand the replies \cite{gundhiCanClassicalTheories2026, biagioSimpleReasonWhy2026, diosiNoClassicalGravity2026}. The paper is organised as follows. In Section~\ref{sec:framework} we summarise the AH framework and isolate the fourth-order term responsible for their claimed entanglement.  In Section~\ref{sec:difnumber} we show that the term depends on the field identity and particle content of the two interferometers, vanishing when the required matter-field cross-contractions are forbidden.  In Section~\ref{sec:integral} we analyse the propagator and long-time approximation, arguing that the $1/r$ contribution corresponds to an on-shell matter-propagation channel rather than a purely virtual gravitational mediation.  In Section~\ref{sec: postselection} we discuss the role of the projection onto the BMV final-state subspace, disputing the non-separability claims of the term, in the lines of \cite{gundhiCanClassicalTheories2026}.  In Section~\ref{sec:spin} we comment on how internal spin or collective degrees of freedom can further distinguish the systems and expose the matter-sector nature of the mechanism.  We conclude in Section~\ref{sec:discussion} by explaining why the AH effect, even if present within their model, should be understood as contamination rather than as a counterexample to the BMV witness.

\begin{figure}[t]
    \centering
    \includegraphics[width=0.98\linewidth]{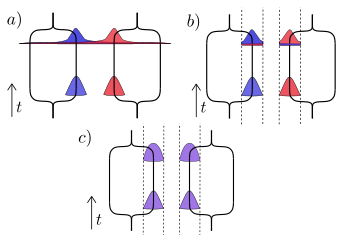}
    \caption{Schematic interpretation of the fourth-order Aziz–Howl matter-sector contribution. (a) A localised matter wavepacket associated with one interferometer develops a small coherent propagation amplitude into the spatial region of the other interferometer, and vice versa. (b) The subsequent projection onto the intended BMV output subspace retains only components supported in the original localised branch modes, thereby hiding the matter-field exchange/leakage sectors. The dashed lines indicate the projected localisation regions. (c) For identical particles, the two crossed propagation histories are indistinguishable and must be added coherently, producing the non-separable branch-dependent amplitude identified by Aziz and Howl. The diagram, therefore, represents matter-field cross-talk in a fixed classical background, rather than quantum information carried by classical gravitational degrees of freedom.}
    \label{fig:leak}
\end{figure}

Our conclusion from this work, at a glance, is that the claimed entangling effect of AH results from indistinguishable particles leaking from one interferometer to the other, plus a postselection process.%

\section{Framework and results of AH}
\label{sec:framework}
Let us introduce a simplified speed-up version of the main framework and results of the AH paper, which are contained within their first two sections of their Supplementary Material \cite{azizClassicalTheoriesGravity2025}. We focus on the calculations of entanglement generation by semiclassical gravity. We start with the Lagrangian density of a free scalar field in a classical curved spacetime. The quantum field is denoted by $\hat{\phi}(x)$ and the classical curved spacetime is given by the metric $g_{\mu \nu}(x)$. Now, we consider a classical spacetime that is a perturbation of flat spacetime given by $g_{\mu \nu} =\eta_{\mu \nu} +h_{\mu \nu}$. The perturbation induced by a classical non-relativistic gravitational potential is used. This treatment of weak-field gravitational fields is standard. Such perturbation is given by $h_{\mu \nu}(x)=-2 \Phi(\vec{x}) \delta_{ \mu \nu}$. $\Phi(\vec{x})$ is the classical gravitational potential. It is a fixed function of position proportional to $G$. $G$ is used as the perturbative small constant, which we will keep at first order; thus, we keep $\Phi(\vec{x})$ only at first order.      

Under this perturbative classical spacetime, we consider the Lagrangian density of a complex massive scalar field in QFT in curved spacetime with a minimal coupling.
\begin{gather}
    \hat{\pazocal{L}}= \sqrt{-|g(x)|} \left( g^{\mu \nu}(x) \partial_\mu \hat{\phi}^\dagger(x) \partial_{\nu} \hat{\phi}(x) - m^2 \hat{\phi}^\dagger(x) \hat{\phi}(x) \right)
\end{gather}
This Lagrangian density form is very standard in QFT in curved spacetime, as it is used in deducing Hawking radiation, for example \cite{hawkingParticleCreationBlack1975}. Now, we keep only the terms at first order in perturbation of the Minkowski metric. The metric is diagonal: $g_{\mu \nu}(x)= \text{diag}(1-2\Phi(\vec{x}), -1-2\Phi(\vec{x}), -1-2\Phi(\vec{x}), -1-2\Phi(\vec{x}))$. So its determinant at first order is given by $|g(x)| \approx -1 -4 \Phi(\vec{x})$; therefore the square root of minus the determinant of the metric term will be at first order $\sqrt{-|g(x)|}\approx 1+2 \Phi(\vec{x})$. The inverse of the metric is then given by  $g^{\mu \nu}(x)\approx \text{diag}(1+2\Phi(\vec{x}), -1+2\Phi(\vec{x}), -1+2\Phi(\vec{x}), -1+2\Phi(\vec{x}))$. Writing explicitly and dropping all the terms that are second order or higher, our Lagrangian density is 
\begin{equation}
\begin{aligned}
\hat{\pazocal{L}} \approx{}&
(1+4\Phi(\vec{x}))
\,\partial_0 \hat{\phi}^\dagger(x)\,
\partial_0 \hat{\phi}(x)  \\
&-\sum_{j=1}^3
\partial_j \hat{\phi}^\dagger(x)\,
\partial_j \hat{\phi}(x) \\
&-(1+2\Phi(\vec{x}))m^2
\hat{\phi}^\dagger(x)\hat{\phi}(x).
\end{aligned}
\end{equation}
From this Lagrangian density, we can obtain the Hamiltonian density
\begin{equation}
\begin{aligned}
\hat{\pazocal{H}} ={}&
(1+4\Phi(\vec{x}))
\,\partial_0 \hat{\phi}^\dagger(x)\,
\partial_0 \hat{\phi}(x) \\
&+\sum_{j=1}^3
\partial_j \hat{\phi}^\dagger(x)\,
\partial_j \hat{\phi}(x) \\
&+(1+2\Phi(\vec{x}))m^2
\hat{\phi}^\dagger(x)\hat{\phi}(x).
\end{aligned}
\end{equation}
which can be split between the Hamiltonian of the free field in the flat Minkowski metric $\hat{H}_0$ and an interacting Hamiltonian that arises due to the first-order perturbation of the metric $\hat{H}_{int}$. Using the Minkowski free field Lagrangian density we can find the conjugate fields $\hat{\pi}(x)=\frac{\partial \hat{\pazocal{L}}_0}{\partial \partial_0 \hat{\phi}(x) }=\partial_0 \hat{\phi}^\dagger(x)$ and  $\hat{\pi}^\dagger(x)=\frac{\partial \hat{\pazocal{L}}_0}{\partial \partial_0 \hat{\phi}^\dagger(x) }=\partial_0 \hat{\phi}(x)$. We define the canonical momenta with respect to the Minkowski free Lagrangian and treat all $\Phi$-dependent corrections perturbatively as interactions. We now obtain the expressions of the Hamiltonians
\begin{align}
\hat{H}_0
={}& \int d^3\vec{x}\,
\Bigg[
\hat{\pi}^\dagger\hat{\pi}
+\sum_{j=1}^3
\partial_j\hat{\phi}^\dagger\,
\partial_j\hat{\phi}
+m^2\hat{\phi}^\dagger\hat{\phi}
\Bigg],
\label{eq:H0}
\\[0.5em]
\hat{H}_{\rm int}
={}& 4\int d^3\vec{x}\,
\Phi(\vec{x})
\Bigg[
\hat{\pi}^\dagger\hat{\pi}
+\frac{m^2}{2}
\hat{\phi}^\dagger\hat{\phi}
\Bigg],
\label{eq:Hint}
\end{align}

The free field Hamiltonian and field decomposition are known, thus allowing us to treat $\hat{H}_{int}$ as an interaction Hamiltonian in the perturbative interaction picture in QFT. In the classical-gravity model, the gravitational field enters as a c-number spacetime-dependent coupling multiplying quantum matter operators.  It does not itself supply non-commuting gravitational degrees of freedom. In the interaction picture, to obtain the usual Dyson series that gives rise to the famous Feynman diagrams in perturbative QFT, we need the Hamiltonian in the interaction picture $\hat{H}_I(t)=e^{i \hat{H_0} t} \hat{H}_{int} e^{-i \hat{H}_0 t}$. $\hat{H}_I(t)$ is just $\hat{H}_{int}$ in the Heisenberg picture, where the field observables are time-dependent. Concretely, we have
\begin{equation}
\begin{aligned}
\hat{\phi}(x)
&\equiv \hat{\phi}(t,\vec{x})  \\
&= e^{i\hat{H}_0 t}
   \hat{\phi}(0,\vec{x})
   e^{-i\hat{H}_0 t}  \\
&= \int \frac{d^3\vec{k}}{(2\pi)^3}
   \frac{1}{\sqrt{2\omega_{\vec{k}}}}
   \left(
   e^{-ikx}\hat{a}_{\vec{k}}
   + e^{ikx}\hat{b}_{\vec{k}}^\dagger
   \right).
\end{aligned}
\end{equation}
with $k_0=\omega_{\vec{k}}=\sqrt{|\vec{k}|^2+m^2}$ in the exponential. 

The authors of AH \cite{azizClassicalTheoriesGravity2025} use the perturbative approach to know features of the evolution of an initial state $\ket{\psi_0}$. The time-evolved state corresponds to the initial state having elapsed a finite time $t$ in the curved spacetime given by the perturbation of the Minkowski metric, the potential $\Phi(\vec{x})$. The authors consider the fixed potential $\Phi(\vec{x})$ to be sourced by the initial matter state $\ket{\psi_0}$, as is usual in semiclassical gravity. The initial state of interest for modelling the standard BMV scenario consists of two massive atoms in two interferometers that gravitationally influence each other. To model this, AH use the states $\ket{N,0}_{1 i} \ket{N,0}_{2 j}$. These are $2N$ particle states where each interferometer has a collection of $N$ particles localised together in a sphere of radius $R<< d_{ij}$, and no antiparticles are present. Concretely, we have that $\ket{N,0}_{1i}\ket{N,0}_{2j}
=$
\begin{equation}
\begin{aligned}
=&
\prod_{k=1}^{N}
\Bigg[
\int d^3\vec{x}_k\,
\tilde{\phi}_{1i}(\vec{x}_k)
\int d^3\vec{p}_k\,
e^{-i\vec{p}_k\cdot\vec{x}_k}
\hat{a}_{\vec{p}_k}^\dagger
\Bigg]
\\
&\times
\prod_{k=1}^{N}
\Bigg[
\int d^3\vec{y}_k\,
\tilde{\phi}_{2j}(\vec{y}_k)
\int d^3\vec{l}_k\,
e^{-i\vec{l}_k\cdot\vec{y}_k}
\hat{a}_{\vec{l}_k}^\dagger
\Bigg]
\ket{0}.
\end{aligned}
\end{equation}
where $\tilde{\phi}_{\kappa i}(\vec{x})=\theta(R-|\vec{x}-\vec{X}_{\kappa i}|)/\sqrt{V}$ is the uniform single-particle wavefunction that localises a single particle within a sphere of radius $R$ centered at the point $\vec{X}_{\kappa i}$. So, these states are products of single-particle states with $N$ uniformly localised at the branch $i$ of the interferometer 1 and $N$ localised at the branch $j$ of the interferometer 2. 

Since in the BMV experiment it is natural to consider a uniform superposition at each of the interferometers, the initial state considered is the following
\begin{equation}
\begin{aligned}
\ket{\psi_0}
=&\frac{1}{2}\Big(
\ket{N,0}_{1L}\ket{N,0}_{2L}
+\ket{N,0}_{1R}\ket{N,0}_{2L}
\\
&+\ket{N,0}_{1L}\ket{N,0}_{2R}
+\ket{N,0}_{1R}\ket{N,0}_{2R}
\Big).
\end{aligned}
\end{equation}

The authors consider the evolution of this initial state to interrogate whether it has become entangled. The time evolution after some finite time $t$ is given by 
\begin{gather}
    \ket{\psi(t)}=e^{i \hat{H}_0 t} \hat{\pazocal{T}} e^{-i \int_0^t d\tau \hat{H}_I(\tau) } \ket{\psi_0}
\end{gather}
It is very difficult to compute the evolution exactly. The authors use the Dyson series expansion of the interaction unitary involving $\hat{H}_I(\tau)$ to assess the evolution perturbatively. Since each $\hat{H}_I$ is a first-order term due to the presence of $\Phi(\vec{x})$, the authors compute contributions up to the $ 4$th order, and claim their results showcase that $\ket{\psi(t)}$ is entangled. 

To calculate these terms, they use the Wick contraction theorem. To use it, one needs to evaluate transition amplitudes, not states, so Aziz and Howl calculate the transition amplitudes from the initial state to the general form of state they expect after the time evolution. They reason that since the states involved in the initial state are localised states that under the free Hamiltonian are eigenstates when considering the non-relativistic approximation, the generic final state should also be a superposition of the freely evolved initial states under the non-relativistic approximation. Explicitly, they assume that the final state is of the following form:
\begin{equation}
\begin{aligned}
\ket{\psi_f}&=
\alpha_{LL}\ket{N,0}_{1L}\ket{N,0}_{2L}
+\alpha_{LR}\ket{N,0}_{1L}\ket{N,0}_{2R}
\\
&+\alpha_{RL}\ket{N,0}_{1R}\ket{N,0}_{2L}
+\alpha_{RR}\ket{N,0}_{1R}\ket{N,0}_{2R}
\end{aligned}
\end{equation}
In our view, this is a reasonable procedure, since we expect the states to remain unaltered spatially, and we want to consider only which phases they accumulate. 

However, one needs to be careful with the conclusions we can draw from this analysis because we are guessing the final state and essentially preventing other terms from being superposed in it. In other words, this should be understood as a projection of the full QFT evolution onto the BMV final-state subspace, rather than as a derivation of the most general final state. As we will discuss, this spatial truncation is particularly relevant when assessing the presence of entanglement in the final state.%

\begin{figure}[!ht]
    \centering
    \vspace{12pt}
    \includegraphics[width=0.65\linewidth]{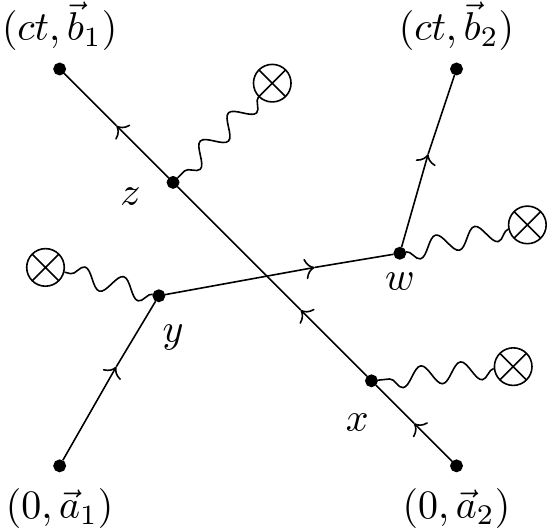}
    \caption{The Feynman diagram representing the entangling term. The directed lines correspond to particles. The black dots are vertices. The vertices with the wavy line correspond to terms of the interacting Hamiltonian, where the wavy line corresponds to the fixed global position function $\Phi(\vec{x})$.}
    \label{fig:ent}
\end{figure}
The physical reasoning behind the assumption that the final state is of this form is that the gravitational interaction being modelled will not generate diffusion of the localised wavepackets in the interferometers' arms. Through this truncation of the final state, we are \emph{de facto} preventing the final state from including terms that would be explained by the diffusion and spatial shift of the one-particle wavepackets that form the $2N$ particle state. One is cutting off terms in which a single particle has evolved out of the localisation region of the branches in the interferometers, even by as little as an $\varepsilon$ distance. The truncation procedure also forbids the change in shape in the uniformly localised wavepackets in each interferometer arm. We believe these are reasonable assumptions given the mental idea we have of the BMV experiment, where we know in the linearised gravity case \cite{boseSpinEntanglementWitness2017,azizClassicalTheoriesGravity2025} that the leading non-relativistic terms imply that the gravitational interaction would only contribute different phases to the matter in the two interferometers' arms. However, we are in a different scenario with semiclassical gravity and a different level of approximation in the non-relativistic limit, so these assumptions might not be justified or implemented correctly. Despite its reasonableness, we need to keep in mind, throughout the work, the motivations and reasoning behind the truncation when analysing the conclusions of the AH claim.     

Having done the spatial truncation, the authors compute the transition amplitudes $\alpha_{ij}$, Eq. 14 of AH Supplementary Material \cite{azizSupplementaryClassicalTheoriesGravity2025a}: 
\begin{gather}
    \alpha_{ij}=\bra{N,0}_{1 i} \bra{N,0}_{2 j} e^{i \hat{H}_0 t} \hat{\pazocal{T}} e^{-i \int_0^t d\tau \hat{H}_I(\tau)} \ket{\psi_0}
\end{gather}

A point that will be important below is that the replacement of the full initial superposition by the corresponding same-branch component is stronger than ordinary orthogonality of the branch states.  Orthogonality gives
\begin{equation}
{}_{1i}\!\bra{N,0}{}_{2j}\!\bra{N,0}\cdot \ket{N,0}_{1k}\ket{N,0}_{2l}
\propto \delta_{ik}\delta_{jl},
\end{equation}
but it does not by itself imply that
\begin{equation}
{}_{1i}\!\bra{N,0}{}_{2j}\!\bra{N,0}\hat U_I
\ket{N,0}_{1k}\ket{N,0}_{2l}=0
\end{equation}
for \((k,l)\neq(i,j)\).  This additional step is harmless only if the dynamics preserve the four branch sectors.  The fourth-order AH diagram is precisely the place where this assumption becomes nontrivial, because it contains matter-field propagation between the localised regions.

The free evolution can be ignored because in the non-relativistic limit, it only contributes a phase to $\bra{N,0}_{1 i} \bra{N,0}_{2 j}$ which does not depend on which interferometer branches $i,j$ we are considering. Using the Dyson series expansion of the interaction unitary, we obtain the order contributions to $\alpha_{ij}= \sum_{n=0}^\infty \alpha_{ij}^{(n)}$, where each term is given by 
\begin{gather}
\alpha_{ij}^{(n)}
={}
\frac{(-i)^n}{n!}\,
{}_{1i}\!\bra{N,0}\,
{}_{2j}\!\bra{N,0}\,
\hat{T}
\int_0^t d\tau_1 \cdots \int_0^t d\tau_n
\nonumber \\
\times
\hat{H}_I(\tau_1)\cdots \hat{H}_I(\tau_n)
\ket{\psi_0}.
\end{gather}

To evaluate these transition amplitudes, the Wick contraction theorem is applied. Feynman diagrams are used to visualise the different contributions being calculated. The entangling term that AH claim appears at fourth order within $\alpha^{(4)}_{ij}$. Concretely, the entangling term is the non-separable contribution $\beta^{(4)}_{ij}$ in AH's work. 

We showcase in Figure \ref{fig:ent} the Feynman diagrammatic representation of this term. There are other fourth-order associated Feynman diagrams in which the propagators do not cross from one interferometer to the other. Still, these are said to contribute only relative phases between the two interferometers. Thus, the Feynman diagram above is the one at the lowest order that is claimed to generate an entangling term between the interferometers.

 \begin{widetext}
The expression for the AH entangling term $\beta_{ij}^{(4)}$ in equation form is:
\begin{gather}
    \beta^{(4)}_{i j}=\frac{1}{16}\int_I d^4 x \int_I d^4 y \int_I d^4 z \int_I d^4 w h^{\mu \nu }(w)  h^{\rho \sigma }(z) h^{\gamma \delta }(y) h^{\kappa \lambda }(x) \times \label{eq:beta} \\ \times \wick{
    \langle \c1{N,0} \vert_{1 i} \langle \c2{N,0} \vert_{2 j} \hat{\pazocal{T}}_{ \mu \nu }[ \c2{\hat{ \phi}^\dagger}(w) \c3{\hat{\phi}}(w) ]  \hat{\pazocal{T}}_{ \rho \sigma }[ \c1{\hat{\phi}^\dagger}(z) \c4{\hat{\phi}}(z)] \hat{\pazocal{T}}_{ \gamma \delta }[ \c3{\hat{\phi}^\dagger}(y) \c5{\hat{\phi}}(y)] \hat{\pazocal{T}}_{ \kappa \lambda }[ \c4{\hat{\phi}^\dagger}(x) \c2{\hat{\phi}}(x)] \vert \c5{N,0}\rangle_{1 i} \vert \c2{N,0} \rangle_{2 j} 
    } \nonumber 
\end{gather}
\end{widetext}
where $h^{\mu \nu }(x) \hat{\pazocal{T}}_{ \mu \nu}[ \hat{\phi}^\dagger(x) \hat{\phi}(x)]=$
\begin{equation}
     4 \Phi(\vec{x}) \left(\hat{\pi}^\dagger(x) \hat{\pi}(x) +\frac{m^2}{2} \hat{\phi}^\dagger(x) \hat{\phi}(x)\right).
\end{equation}
and where we use the notation $\int_I d^4 x := \int_0^t dx_0 \int d^3 \vec{x}$.

Having introduced the basis formalism that AH use in detail \cite{azizClassicalTheoriesGravity2025}, we are well suited to understand and analyse their claims directly, not relying on other methods of computation or arguments.     

\section{No entanglement of different particle type}\label{sec:difnumber}
The first point we raise in our analysis of the AH result is that the proposed entangling mechanism is highly non-generic. The mechanism described by AH produces the claimed non-separable fourth-order contribution only when the two interferometers are modelled as compatible excitations of the same quantum matter field, so that the required cross-contractions of the matter operators are non-vanishing. If, instead, the initial state $\ket{\psi_0}$ were replaced by
\begin{equation}
\begin{aligned}
\ket{\eta_0}
=\frac{1}{2}\Big(
&\ket{N,0}_{1L}\ket{0,N}_{2L}
+\ket{N,0}_{1R}\ket{0,N}_{2L}  \\
&+\ket{N,0}_{1L}\ket{0,N}_{2R}
+\ket{N,0}_{1R}\ket{0,N}_{2R}
\Big),
\end{aligned}
\end{equation}
where the object in the first interferometer is composed of particles while the object in the second interferometer is composed of antiparticles, then the crossed fourth-order term identified by AH as entangling vanishes within the same non-relativistic contraction rules. The reason is not that gravity fails to couple to antiparticles. Rather, the matter-field exchange channel required by the AH diagram cannot connect the projected initial and final charge sectors. The corresponding virtual matter-exchange process is therefore absent.

We now demonstrate this explicitly. The entangling contribution arises from the usual Dyson expansion of the perturbative interaction Hamiltonian, as summarised in Sec.~\ref{sec:framework}. As discussed there, the relevant contribution is represented by the Feynman diagram shown in Fig.~\ref{fig:ent}. In this diagram, the external upper legs represent the projected final states of the experiment. We use the notation $\ket{N, M}_{\kappa i}$ to denote a localised state containing $N$ particles and $M$ antiparticles at the $i$th branch of interferometer $\kappa$.

If we instead choose the asymmetric particle--antiparticle initial state $\ket{\eta_0}$, then by the same reasoning used by AH the projected final state would be
\begin{equation}
\begin{aligned}
\ket{\eta_f}
&={}
\xi_{LL}\ket{N,0}_{1L}\ket{0,N}_{2L}
+\xi_{RL}\ket{N,0}_{1R}\ket{0,N}_{2L} \\
&+\xi_{LR}\ket{N,0}_{1L}\ket{0,N}_{2R}
+\xi_{RR}\ket{N,0}_{1R}\ket{0,N}_{2R}.
\end{aligned}
\end{equation}

\begin{figure}[!hb]
    \centering
    \includegraphics[width=0.65\linewidth]{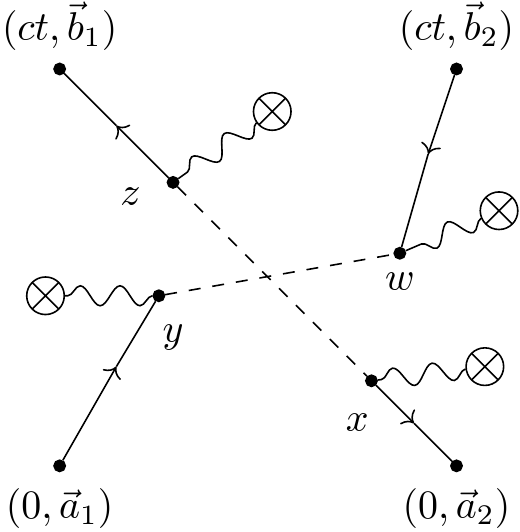}
    \caption{Analogue of the AH fourth-order crossed diagram when the second interferometer is composed of antiparticles. The arrows on the right interferometer are reversed relative to the particle case. The dashed lines indicate the contractions required to reproduce the crossed matter-exchange channel. No oriented particle propagator can replace the dashed lines while maintaining one incoming and one outgoing matter line at each vertex. Consequently, the corresponding Wick contractions vanish.}
    \label{fig:antiparticlediag}
\end{figure}

However, substituting these initial and final states into the relevant AH diagram gives a vanishing amplitude. In the diagram, the initial external leg associated with the first interferometer always contains $N$ particles. This line scatters from the classical gravitational background into a virtual matter propagator, and then scatters again into the external leg associated with the second interferometer. But in the projected final state of the second interferometer, there are $N$ antiparticles, not $N$ particles. The crossed contraction would therefore have to convert a particle line into an antiparticle line, which is forbidden by the non-vanishing contractions of the complex scalar field. This conclusion is independent of how the contractions are arranged. To reproduce the analogue of the AH crossed diagram in Fig.~\ref{fig:antiparticlediag}, where the virtual matter propagators exchange the two interferometers, one must contract the operators at $x$ with those at $z$, and those at $y$ with those at $w$. In Fig.~\ref{fig:antiparticlediag}, $x$ is contracted with the initial right external leg of antiparticles, $y$ with the initial left external leg of particles, $w$ with the final right external leg of antiparticles, and $z$ with the final left external leg of particles. These contractions cannot be completed with non-vanishing particle-number and charge assignments.

As shown in Eqs.~22 and 23 of the Supplementary Material of AH~\cite{azizClassicalTheoriesGravity2025}, the contractions with the localised particle and antiparticle states obey

\begin{align}
&\wick{\c2{\hat{\phi}}(x) |\c2{N},0\rangle_{\kappa i}}
=
\frac{\sqrt{N}}{\sqrt{2m}} e^{imx_0}\tilde{\phi}_{\kappa i}(\vec{x}) \ket{N-1,0}_{\kappa i} \label{eq:contractket} \\
&\wick{\c2{\hat{\phi}^\dagger}(x)|\c2{N},0\rangle_{\kappa i}}
=0
\\
&\wick{\c2{\hat{\phi}^\dagger}(x)|0,\c2{N}\rangle_{\kappa i}}
=
\frac{\sqrt{N}}{\sqrt{2m}} e^{imx_0}\tilde{\phi}_{\kappa i}(\vec{x})\ket{0,N-1}_{\kappa i}
\\
&\wick{\c2{\hat{\phi}}(x) |0,\c2{N}\rangle_{\kappa i}}
=0
\end{align}

The corresponding contractions involving $\hat{\pi}(x)$ and $\hat{\pi}^\dagger(x)$ have the same selection rules, up to different proportionality constants and the corresponding interchange of daggered and undaggered operators. The anomalous contractions vanish:
\begin{gather}
\wick{\c2{\hat{\phi}}(x)\c2{\hat{\phi}}(y)}
=
\wick{\c2{\hat{\phi}^\dagger}(x)\c2{\hat{\phi}^\dagger}(y)}
=
\wick{\c2{\hat{\pi}}(x)\c2{\hat{\pi}}(y)}
=
\wick{\c2{\hat{\pi}^\dagger}(x)\c2{\hat{\pi}^\dagger}(y)}=
\nonumber \\
=
\wick{\c2{\hat{\pi}}(x)\c2{\hat{\phi}^\dagger}(y)}
=
\wick{\c2{\hat{\pi}^\dagger}(x)\c2{\hat{\phi}}(y)}
=0 .
\end{gather}

We can now apply these rules to the diagram involving antiparticle states. Consider first the terms in which all four interaction Hamiltonians contribute $\hat{\phi}^\dagger\hat{\phi}$ bilinears. 

Moving from the external states inward, we select the only choices that do not immediately force the expression to vanish. We contract $\hat{\phi}^\dagger(x)$ with $\ket{0,N}_{2l}$, $\hat{\phi}(y)$ to $\ket{N,0}_{1k}$, $\bra{0,N}_{2j}$ with $\hat{\phi}(w)$ and $\bra{N,0}_{1i}$ with $\hat{\phi}^\dagger(z)$.
\begin{widetext}
\begin{equation}
\begin{aligned}
\wick{
{}_{1i}\!\langle\c3{N},0|\,
{}_{2j}\!\langle \c2{0},N|\,
\hat{\phi}^\dagger(w)\c2{\hat{\phi}}(w)
\c3{\hat{\phi}^\dagger}(z)\hat{\phi}(z)
\hat{\phi}^\dagger(y)\c3{\hat{\phi}}(y)
\c2{\hat{\phi}^\dagger}(x)\hat{\phi}(x)
|\c3{N,0}\rangle_{1k}
|\c2{0,N}\rangle_{2l}
}.
\end{aligned}
\end{equation}
\end{widetext}
To obtain the crossed AH matter-exchange channel, the remaining operators would have to be contracted between $y$ and $w$, and between $x$ and $z$. These are contractions of the form $\hat{\phi}^\dagger(y)$ with $\hat{\phi}^\dagger(w)$, and $\hat{\phi}(x)$ with $\hat{\phi}(z)$, both of which vanish. The same conclusion follows for the terms involving $\hat{\pi}^\dagger\hat{\pi}$ in the interaction Hamiltonian, since they obey the same particle-number and charge-selection rules.

The physical interpretation is straightforward. The AH crossed contribution corresponds to a matter-sector exchange amplitude in which a particle associated with one interferometer branch propagates into the other interferometer, with the reciprocal process occurring in the opposite direction. If the second interferometer instead contains $N$ antiparticles, then the propagation of a particle from the first interferometer into the second would yield a state of the form $\ket{1, N-1}_{2j}$. Such a state is not contained in the projected final-state ansatz. It has been excluded by the same non-relativistic truncation used to restrict the analysis to the expected BMV output subspace. Thus, within this projected subspace, the AH crossed contribution cannot generate entanglement for this asymmetric particle--antiparticle configuration.

At this point, we arrive at a fork in the interpretation. The claim that the final-state truncation is sufficiently justified to allow gravity to entangle two collections of $N$ particles of the same type does not come for free. The resulting branch-dependent term is sensitive not only to the masses and separations of the two systems, but also to whether they belong to compatible quantum matter sectors. This behaviour is not expected of a universal gravitational phase. Rather, it indicates that the matter field carries the non-separable contribution. When the particle sectors are incompatible, the AH fourth-order classical-gravity mechanism does not produce entanglement.

This conclusion is striking. For genuine entanglement generated by the gravitational interaction between two systems, one expects the relevant parameters to be their masses, relative separations, velocities, interaction time, and possibly the classical gravitational background. Internal degrees of freedom such as spin may also enter in more refined models. However, one would not expect the microscopic particle type to determine whether gravity can generate entanglement. It would be surprising if a purely gravitational interaction could distinguish, in this sense, whether one system consisted of electrons or positrons, hydrogen or antihydrogen, or other such matter-sector choices.

As we discuss in detail in the following sections, we believe that this final-state truncation is a source of conceptual confusion and must be treated with care. In this section, we have highlighted a simple consequence of the AH mechanism: the claimed entangling contribution depends on the compatibility, and in particular the indistinguishability, of the underlying matter-field excitations in the two interferometers. Identifying this dependence shows that the AH effect is not a universal gravitational entangling channel, but a matter-sector contribution whose interpretation requires careful analysis.

\section{Interpreting the AH phase} \label{sec:integral}

We now turn to the estimate of the AH phase, i.e.\ the fourth-order contribution $\beta^{(4)}_{ij}$ of Eq.~\eqref{eq:beta}. The stress--energy tensor contributes two structures, $\hat{\pi}^\dagger\hat{\pi}$ and $\hat{\phi}^\dagger\hat{\phi}$. We isolate the $\hat{\phi}^\dagger\hat{\phi}$ part as a diagnostic of the spacetime structure of the term; the $\hat{\pi}^\dagger\hat{\pi}$ part must be retained for the final numerical coefficient, but in the non-relativistic contractions it generates the same matter-propagator channel, differing only by powers of $m$ and convention-dependent signs. 

\begin{figure}[!hb]
    \centering
    \includegraphics[width=0.47\linewidth]{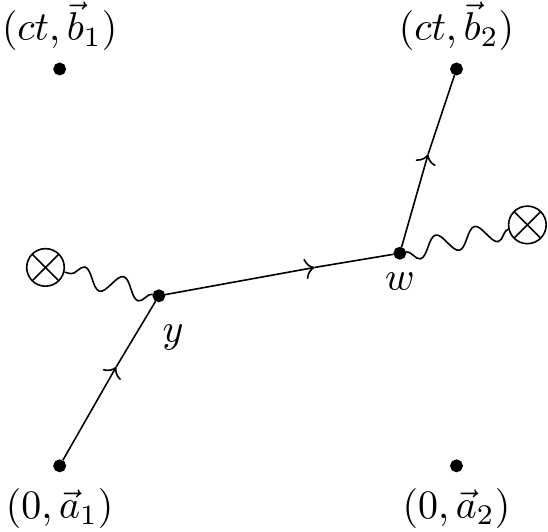} \quad
    \includegraphics[width=0.47\linewidth]{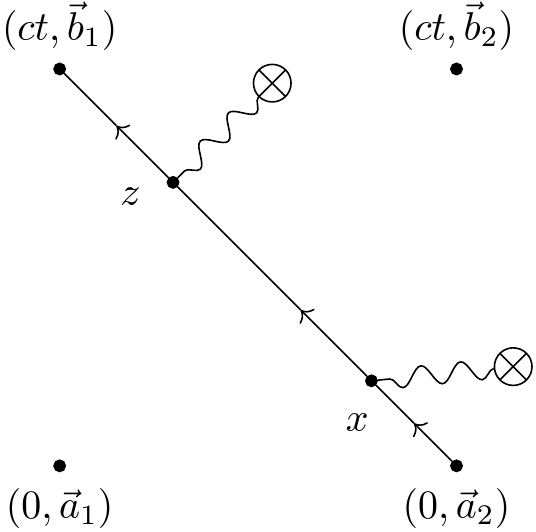}
    \caption{Factorisation of the selected fourth-order Wick contraction into two identical matter-propagator subamplitudes. Each subamplitude connects one localised branch mode to the other through the quantum matter field in the fixed classical gravitational background.}
    \label{fig:product}
\end{figure}

Because the relevant field contractions are ordinary Feynman propagators, the amplitude factorises into two identical matter-propagator subamplitudes, each connecting one localised branch mode to the other through the quantum matter field in the fixed classical background, as showcased in  (Fig.~\ref{fig:product}).%

Explicitly, contracting each field operator with the localised $N$-particle states leaves a localised $(N-1)$-particle state multiplied by local wavefunction factors, so that%
\begin{widetext}
\begin{gather}
    \beta_{ij}^{(4),\phi,\phi}= \frac{m^4}{4}\int_I d^4 x \int_I d^4 y \int_I d^4 z \int_I d^4 w \Phi(\vec{w})  \Phi(\vec{z}) \Phi(\vec{y}) \Phi(\vec{x}) \times  \\ \times \wick{
    \langle \c1{N,0} \vert_{1 i} \langle \c2{N,0} \vert_{2 j}  \c2{\hat{ \phi}^\dagger}(w) \c3{\hat{\phi}}(w) \c1{\hat{\phi}^\dagger}(z) \c4{\hat{\phi}}(z) \c3{\hat{\phi}^\dagger}(y) \c5{\hat{\phi}}(y) \c4{\hat{\phi}^\dagger}(x) \c2{\hat{\phi}}(x) \vert \c5{N,0}\rangle_{1 i} \vert \c2{N,0} \rangle_{2 j} 
    } \nonumber= \\=\frac{N^2 m^2}{16}\int d^3 x \int d^3 y \int d^3 z \int d^3 w \Phi(\vec{w})  \Phi(\vec{z}) \Phi(\vec{y}) \Phi(\vec{x})   \tilde{\phi}_{2 j}(\vec{x} ) \tilde{\phi}_{1 i}(\vec{z} )  \tilde{\phi}_{2 j}(\vec{w} ) \tilde{\phi}_{1 i}(\vec{y} ) \times \nonumber \\ \times \int_0^t d x_0  \int_0^t d y_0  \int_0^t d z_0  \int_0^t d w_0e^{i m (w_0-y_0)} e^{i m (z_0-x_0)} D_F(x-z) D_F(y-w) = \\ = \frac{N^2 m^2}{16}\left(\int d^3 x  \int d^3 z    \Phi(\vec{z}) \Phi(\vec{x})   \tilde{\phi}_{2 j}(\vec{x} ) \tilde{\phi}_{1 i}(\vec{z} )   \int_0^t d x_0   \int_0^t d z_0  e^{i m (z_0-x_0)} D_F(x-z) \right) \times \nonumber \\ \left(\int d^3 y  \int d^3 w    \Phi(\vec{w}) \Phi(\vec{y})   \tilde{\phi}_{2 j}(\vec{w} ) \tilde{\phi}_{1 i}(\vec{y} )   \int_0^t d y_0   \int_0^t d w_0  e^{i m (w_0-y_0)} D_F(y-w) \right) 
\end{gather}
\end{widetext}

These two independent terms contain the same double time integral $I(\vec{x}-\vec{z})$
\begin{gather}
    I(\vec{x}-\vec{z})=  \int_0^t d x_0   \int_0^t d z_0  e^{i m (z_0-x_0)} D_F(x-z)
\end{gather}
where the Feynman propagator is 
\begin{gather}
D_F(x-z)=\int \frac{d^4 p}{(2\pi)^4} \frac{i }{p^2-m^2+i \varepsilon} e^{-i p \cdot (x-z)}    
\end{gather}
where we use the $(+---)$ signature, so that $p^2=p_0^2-|\vec p|^2$. 

We evaluate the double time integral by first carrying out the spatial momentum integral in spherical coordinates.

\begin{gather}
    J=\int d^3 \vec{p} \frac{i}{p_0^2-|\vec{p}|^2-m^2+i\varepsilon}  e^{i \vec{p}\cdot (\vec{x}-\vec{z})}  \end{gather} 
\begin{widetext}
\begin{gather}
J =(2 \pi) i e^{-i p_0(x_0-z_0)} \int_0^\infty dp \frac{p^2}{p_0^2-p^2-m^2+i\varepsilon } \int_{-1}^1 d \eta e^{i p |\vec{x}-\vec{z}| \eta} =\nonumber \\ = \frac{(4 \pi) i}{|\vec{x}-\vec{z}|} e^{-i p_0(x_0-z_0)} \int_{-\infty}^\infty dp \frac{p}{p_0^2-p^2-m^2+i\varepsilon }e^{i p |\vec{x}-\vec{z}| }=\nonumber \\ = \frac{-4 \pi i}{|\vec{x}-\vec{z}|} e^{-i p_0(x_0-z_0)} \int_{-\infty}^\infty dp \frac{p}{\left(p-\sqrt{p_0^2-m^2+i\varepsilon}\right) \left(\sqrt{p_0^2-m^2+i\varepsilon}+p\right) }e^{i p |\vec{x}-\vec{z}| }=\nonumber \\ =\frac{-2 \pi i}{|\vec{x}-\vec{z}|} e^{-i p_0(x_0-z_0)} \left(\int_{-\infty}^\infty dp \frac{1}{p-\sqrt{p_0^2-m^2+i\varepsilon} }e^{i p |\vec{x}-\vec{z}| } + \int_{-\infty}^\infty dp \frac{1}{p+\sqrt{p_0^2-m^2+i\varepsilon} }e^{i p |\vec{x}-\vec{z}| } \right)
\end{gather}
\end{widetext}
Both integrals have simple poles off the real axis. Because the exponential is $e^{ip|\vec x-\vec z|}$, the arc in the upper half-plane vanishes as $R\to\infty$, so each integral equals $2\pi i$ times the residue of any pole it encloses in the upper half-plane, and zero otherwise. The pole at $+\sqrt{p_0^2-m^2+i\varepsilon}$ lies in the first quadrant, hence in the upper half-plane, while the pole at $-\sqrt{p_0^2-m^2+i\varepsilon}$ lies in the third quadrant, in the lower half-plane. Only the former contributes, and the residue theorem gives  
\begin{gather}
    = \frac{(2 \pi)^2}{|\vec{x}-\vec{z}|} e^{-i p_0(x_0-z_0)}   e^{i \sqrt{p_0^2-m^2+i\varepsilon} |\vec{x}-\vec{z}| } 
\end{gather}
Thus, we obtain the expression for the propagator $D_F(x-z)=$
\begin{gather}
    \frac{1}{|\vec{x}-\vec{z}|} \int_{-\infty}^{\infty} \frac{d p_0}{(2\pi)^2} e^{-i p_0(x_0-z_0)}  e^{i \sqrt{p_0^2-m^2+i\varepsilon} |\vec{x}-\vec{z}| } 
\end{gather}
The integrand is analytic on the real axis: the branch points are displaced from it by the Feynman prescription $i\varepsilon$, leaving only off-axis branch cuts to track. We can therefore re-express the integral of interest as $I(\vec{x}-\vec{z})= $
\begin{gather}
     \int_0^t d x_0   \int_0^t d z_0  e^{-i m (z_0-x_0)} \frac{1}{|\vec{x}-\vec{z}|} \times \nonumber \\ \int_{-\infty}^{\infty} \frac{d p_0}{(2\pi)^2} e^{-i p_0(x_0-z_0)}  e^{i \sqrt{p_0^2-m^2+i\varepsilon} |\vec{x}-\vec{z}| } 
\end{gather}
Our sign conventions follow Peskin and Schroeder~\cite{peskinIntroductionQuantumField1995} and differ from those of AH, which accounts for the sign of the prefactor. The contraction of the field with the localised states is taken in the non-relativistic regime: of the two terms in $\hat\phi(x)$ only the positive-frequency part acts on the $N$-particle state, and its rapid phase is approximated by the rest-energy factor $e^{-imx_0}$, the kinetic dispersion $|\vec p|^2/2m$ being neglected relative to $m$. This is the step that fixes the external legs on the mass shell, $p_0=m$, and, as we show below, is responsible for the resonance that survives the long-time limit. For finite $t$ the order of integration may be exchanged by Fubini's theorem, since the $p_0$ integral converges to a combination of Bessel functions and the remaining time integral runs over a finite domain. Carrying out the $z_0$ integral gives $I(\vec{x}-\vec{z})=$
\begin{gather}
      \int_0^t d x_0  e^{i m x_0} \frac{2}{|\vec{x}-\vec{z}|} \int_{-\infty}^{\infty} \frac{d p_0}{(2\pi)^2} e^{-i p_0 x_0} e^{i \frac{t}{2} (p_0-m)} \cdot \nonumber \\ \cdot \frac{\sin((p_0-m)\frac{t}{2})}{p_0-m}  e^{i \sqrt{p_0^2-m^2+i\varepsilon} |\vec{x}-\vec{z}| } 
\end{gather}

In the long-time limit the kernel $\sin[(p_0-m)t/2]/(p_0-m)\to\pi\,\delta(p_0-m)$ pins the integral to $p_0=m$, and the regularized integral behaves as
\begin{gather}
    I(\vec{x}-\vec{z})\approx  \int_0^t d x_0  e^{i m x_0} \frac{2}{4\pi |\vec{x}-\vec{z}|}  e^{-i m x_0} e^{i \sqrt{i\varepsilon} |\vec{x}-\vec{z}| }= \nonumber \\ =\frac{ t}{2\pi |\vec{x}-\vec{z}|}e^{i \sqrt{i\varepsilon} |\vec{x}-\vec{z}| }
\end{gather}

In the limit $\varepsilon\to0$ the factor $e^{i\sqrt{i\varepsilon}\,|\vec x-\vec z|}\to1$, leaving $I(\vec x-\vec z)\simeq t/(2\pi|\vec x-\vec z|)$. 

\begin{figure}[!hb]
    \centering
    \includegraphics[width=0.78\linewidth]{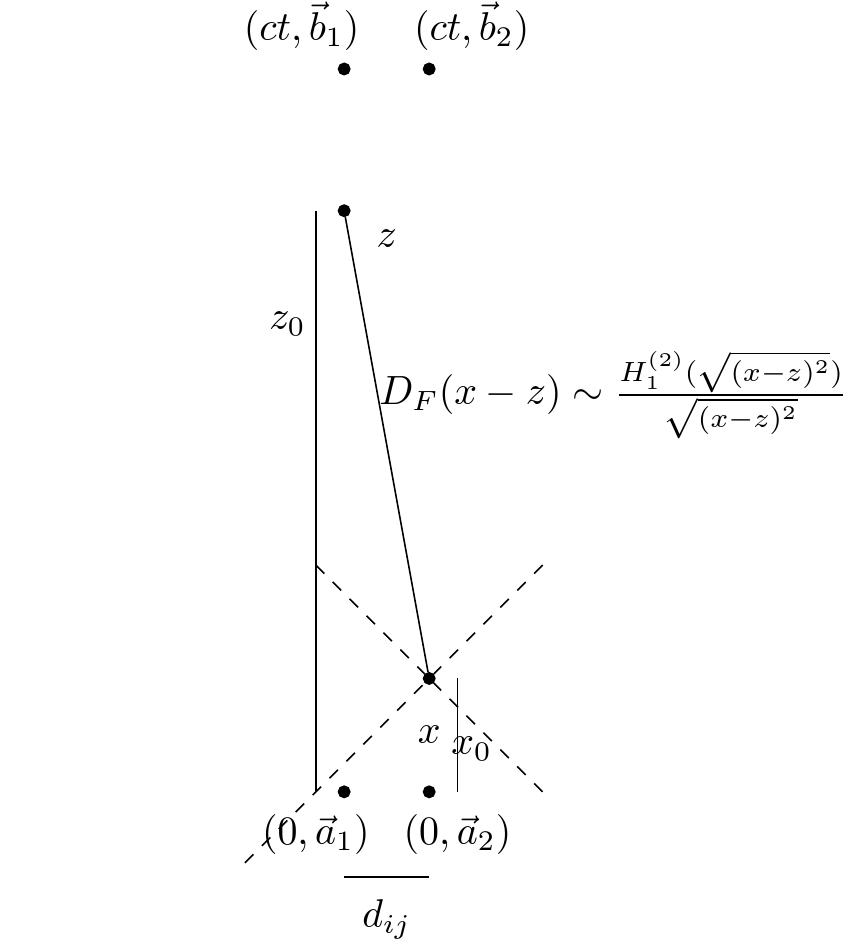}
    \caption{Sketch where we see that because $ct>>> d_{ij}$, the timelike propagator is expected to contribute more in the integral.}
    \label{fig:sketch}
\end{figure}

$\frac{t}{(2\pi|\vec x-\vec z|)}$ is an unsuppressed $1/r$ kernel growing linearly in $t$, in agreement with the AH phase. The two features that control the interpretation below are already visible here — the kernel is purely $1/r$ (no exponential suppression), and it grows linearly with the interaction time.

There is no inconsistency between the finite-$t$ use of Fubini's theorem and the subsequent long-time limit: for every finite $t$ the exchange of integrals is legitimate, and the $t\to\infty$ limit is taken only afterwards, on the resulting expression, where the finite-time kernel $\sin[(p_0-m)t/2]/(p_0-m)$ tends to $\pi\,\delta(p_0-m)$ in the distributional sense. What the limit does change is the physical nature of the contribution: the time-window integral no longer samples a generic off-shell propagator but projects onto the resonant mass-shell component $p_0=m$. The resulting linear growth $I\propto t$ is the hallmark of a real, rate-limited transition (a Fermi golden-rule behaviour) rather than of a bounded off-shell phase, which would saturate at large $t$.

Physically, the surviving contribution describes the real, on-shell propagation of a single matter quantum from one interferometer to the other. The non-relativistic contraction retains only the $e^{-imx_0}$ factor, which is precisely what resonates with the $k_0=m$ pole of the propagator, while the off-shell modes that would carry a genuine virtual exchange average out in the long-time limit. The AH phase is therefore the transition amplitude for one on-shell particle to drift between the two localised modes.

This mechanism does generate entanglement, but of a specific and physically transparent kind. Because real quanta migrate between the interferometers, each carrying a distance-dependent phase, the postselected $N$-particle state acquires correlations between the two arms. The entanglement is thus mediated by matter transport, not by a gravitational degree of freedom. It follows that the effect is exponentially damped by any physical barrier that prevents the free propagation of the non-relativistic quanta between the two interferometers, since such a barrier removes the only channel through which the correlations are built.
 
It is worth addressing directly whether the kernel should instead be exponentially suppressed, as a naive Yukawa intuition would suggest. A massive mediator exchanged \emph{virtually} does produce a Yukawa potential $\sim e^{-mr}/r$, and for atomic masses over laboratory separations the exponent $mr$ is enormous, so such a contribution would be utterly negligible. That suppression, however, is a property of \emph{spacelike} separation: the position-space Feynman propagator of a massive field decays exponentially, as the modified Bessel function $K_1\!\big(m\sqrt{-x^2}\big)$, only when $x$ is spacelike~\cite{peskinIntroductionQuantumField1995}. For \emph{timelike} separation it is governed instead by the Hankel function $H_1^{(2)}\!\big(m\sqrt{x^2}\big)$, which decays as an oscillatory power law and shows no exponential suppression.

In the integral of interest, one has $ct\gg d_{ij}$ (Fig.~\ref{fig:sketch}), so the dominant region of integration is timelike and exponential suppression does not arise. What happens instead is sharper: the $e^{im(x_0-z_0)}$ factor, together with the long-time limit, resonates the $p_0=m$ component and places the exchanged quantum exactly on its mass shell. On-shell $\sqrt{p_0^2-m^2}=0$, the factor $e^{i\sqrt{p_0^2-m^2}\,|\vec x-\vec z|}$ reduces to unity, and one is left with a bare, unsuppressed $1/r$. The absence of Yukawa suppression is therefore not the signature of a long-range gravitational interaction but the signature that the mediating quantum has been forced on-shell — that a \emph{real} particle is being exchanged. A genuinely virtual, off-shell exchange, such as the one obtained at finite $t$ or when the on-shell channel is blocked, remains Yukawa-suppressed and negligible. This is precisely why a physical barrier, which stops real matter transport while leaving any putative gravitational phase untouched, removes the effect.

We stress that our analysis does not dispute the value of the AH phase. We reproduce their estimate and accept it as the correct amplitude for the on-shell channel; our claim is interpretational. That amplitude represents a real matter quantum propagating between the interferometers — a process that is suppressed for distinguishable particles and removed by a physical barrier. It is therefore a matter-sector leakage effect in a classical background, not entanglement mediated by classical gravitational degrees of freedom. 

\section{Is $\beta_{ij}^{(4)}$ separable?}
\label{sec: postselection}

As we have been pointing out throughout the article, the entangling terms of AH correspond to $\beta^{(4)}_{ij}$, which results in the evaluation of the diagram in Figure \ref{fig:ent}. As we have seen already, these terms come from the evaluation in equation form of the contractions: 
\begin{widetext}
\begin{gather}
    \beta^{(4)}_{i j}=\frac{1}{16}\int_I d^4 x \int_I d^4 y \int_I d^4 z \int_I d^4 w h^{\mu \nu }(w)  h^{\rho \sigma }(z) h^{\gamma \delta }(y) h^{\kappa \lambda }(x) \times \\ \times \wick{
    \langle \c1{N,0} \vert_{1 i} \langle \c2{N,0} \vert_{2 j} \hat{\pazocal{T}}_{ \mu \nu }[ \c2{\hat{ \phi}^\dagger}(w) \c3{\hat{\phi}}(w) ]  \hat{\pazocal{T}}_{ \rho \sigma }[ \c1{\hat{\phi}^\dagger}(z) \c4{\hat{\phi}}(z)] \hat{\pazocal{T}}_{ \gamma \delta }[ \c3{\hat{\phi}^\dagger}(y) \c5{\hat{\phi}}(y)] \hat{\pazocal{T}}_{ \kappa \lambda }[ \c4{\hat{\phi}^\dagger}(x) \c2{\hat{\phi}}(x)] \vert \c5{N,0}\rangle_{1 i} \vert \c2{N,0} \rangle_{2 j} 
    } \nonumber 
\end{gather}
\end{widetext}
The terms are entangling according to AH because the transition amplitudes differ for the four branches of the global superposition $i j$. AH claim that the final state is of the form $\ket{\psi_f} \approx \sum_{i,j \in \{L,R\}} \sum_{n=0}^4(\alpha^{(n)}_{ij}) \ket{N,0}_{1i} \ket{N,0}_{2j}$. They claim that, $\alpha_{ij}^{(4)}=\beta^{(4)}_{ij}+...$ and that each $\beta^{(4)}_{i,j}$ term is different. Moreover, by analysing the expression they obtain for $\beta^{(4)}_{ij}$, they conclude that it cannot be expressed as a separable term between the two interferometers as $\beta^{(4)}_{ij}=\varphi_i \varphi_j$. AH shows that, for all terms up to the $4$-th order, such a separable decomposition can be obtained. If the whole state is decomposable in such a way, then there is no entanglement between the two interferometers, but rather local relative phases within them. 

In the Gundhi et al. \cite{gundhiCanClassicalTheories2026} reply to AH, it is claimed that the final state is separable. They show that, at fourth order, there are additional terms that contribute to the final state, giving $\beta_{i j}^{(4)}$ a separable form. They claim that the analysis of AH misses the cross-branch terms associated with the same Feynman diagram used to evaluate $\beta_{ij}^{(4)}$. The terms to refer to are:
\begin{widetext}
\begin{gather}
    \beta^{(4)}_{i j ; k l}=\frac{1}{16}\int_I d^4 x \int_I d^4 y \int_I d^4 z \int_I d^4 w h^{\mu \nu }(w)  h^{\rho \sigma }(z) h^{\gamma \delta }(y) h^{\kappa \lambda }(x) \times \\ \times \wick{
    \langle \c1{N,0} \vert_{1 i} \langle \c2{N,0} \vert_{2 j} \hat{\pazocal{T}}_{ \mu \nu }[ \c2{\hat{ \phi}^\dagger}(w) \c3{\hat{\phi}}(w) ]  \hat{\pazocal{T}}_{ \rho \sigma }[ \c1{\hat{\phi}^\dagger}(z) \c4{\hat{\phi}}(z)] \hat{\pazocal{T}}_{ \gamma \delta }[ \c3{\hat{\phi}^\dagger}(y) \c5{\hat{\phi}}(y)] \hat{\pazocal{T}}_{ \kappa \lambda }[ \c4{\hat{\phi}^\dagger}(x) \c2{\hat{\phi}}(x)] \vert \c5{N,0}\rangle_{1 k} \vert \c2{N,0} \rangle_{2 l} 
    } \nonumber 
\end{gather}

\end{widetext}

where the indices of the branch of the initial state do not necessarily correspond to those of the final state. The four entangling terms of AH are special cases of the full expression, which contains $16$ terms. Gundhi et al. think that dropping these twelve terms is not well justified and that when one includes them, entanglement cannot be claimed. Concretely, we have that $\beta_{ij}^{(4)}=\beta^{(4)}_{ij; ij}$, and the expression for the contribution of this Feynman diagram to the fourth order of the Dyson series of the transition amplitudes should be $\alpha_{ij}^{4}=\frac{1}{2}\sum_{k l} \beta^{(4)}_{ij; kl}+...$. Gundhi et al. justify why this is the correct contribution and why it is not justified to drop the twelve missing terms. Then, by analysing the evaluation of the diagram, following AH procedures, Gundhi et al. can see that the terms $\beta^{(4)}_{i j; k l}$ are separable as $\beta^{(4)}_{i j k l}=\theta_{i l} \eta_{j k} $.  

The reasoning why these terms are not considered in AH is in the comments on Equation 14 of the Supplementary Material \cite{azizSupplementaryClassicalTheoriesGravity2025a}. The authors considered a general tentative final state because under free evolution in the non-relativistic limit, the localised states of the field at the four branches of the superposition are eigenstates of the free evolution. Thus, the proposed generic final state is a general superposition of $\ket{N,0}_{1 i} \ket{N,0}_{2 j}$. Remember that the initial state is $\ket{\psi_0}=\frac{1}{2} \sum_{ij} \ket{N,0}_{1 i} \ket{N,0}_{2 j}$, and the general final state is $\ket{\psi_0}=\sum_{ij} \alpha_{ij} \ket{N,0}_{1 i} \ket{N,0}_{2 j}$. Thus, what we should evaluate to find $\alpha_{ij}$ is 

\begin{gather}
    \alpha_{ij}= \bra{N,0}_{1 i} \bra{N,0}_{2 j} e^{i \hat{H}_0} e^{-i \int_0^t d\tau \hat{H}_I(\tau)} \ket{\psi_0}
\end{gather}
 
\begin{widetext}
   So, in general, by plugging in the expression of the initial state, we obtain
   \begin{gather}
\alpha_{ij}=\bra{N,0}_{1 i} \bra{N,0}_{2 j} e^{i \hat{H}_0} e^{-i \int_0^t d\tau \hat{H}_I(\tau)} \left(\frac{1}{2}\sum_{kl} \ket{N,0}_{1k} \ket{N,0}_{2 l} \right)= \nonumber \\=\frac{1}{2}\sum_{kl} \bra{N,0}_{1 i} \bra{N,0}_{2 j} e^{i \hat{H}_0} e^{-i \int_0^t d\tau \hat{H}_I(\tau)} \ket{N,0}_{1k} \ket{N,0}_{2 l} \label{eq:actualalpha}
\end{gather} 
\end{widetext}
However, AH does not obtain such an expression. Instead, they enforce that the contributions to the final state $\alpha_{ij}$ should be only from the branch in the initial state with paths in $i,j$ as well. So, in the paper, they strictly enforce that the product states of the particles being in branch $i$ in interferometer 1, and in branch $j$ in interferometer 2, are necessarily eigenstates of the interaction unitary. Such a claim is much stronger than claiming that the final state is expected to be a superposition of the four localised product states of the particles in each interferometer arm. We believe it is not justified to drop such terms if they are of the same order as the original entangling terms. From our perspective, at this order, it is not justified to expect the particles to remain in their original branch, since we have seen the term $\beta^{(4)}_{ij}$ to precisely be the scattering of real particles from one interferometer to the other, due to the resonance in the on-shell terms.  
 
\begin{widetext}
The expression for $\beta^{(4)}_{ij ; kl}$ Gundhi et al. give is the following:
\begin{gather}
    \beta^{(4)}_{i j; k l}=\frac{1}{16}\int_I d^4 x \int_I d^4 y \int_I d^4 z \int_I d^4 w h^{\mu \nu }(w)  h^{\rho \sigma }(z) h^{\gamma \delta }(y) h^{\kappa \lambda }(x) \times \\ \times \wick{
    \langle \c1{N,0} \vert_{1 i} \langle \c2{N,0} \vert_{2 j} \hat{\pazocal{T}}_{ \mu \nu }[ \c2{\hat{ \phi}^\dagger}(w) \c3{\hat{\phi}}(w) ]  \hat{\pazocal{T}}_{ \rho \sigma }[ \c1{\hat{\phi}^\dagger}(z) \c4{\hat{\phi}}(z)] \hat{\pazocal{T}}_{ \gamma \delta }[ \c3{\hat{\phi}^\dagger}(y) \c5{\hat{\phi}}(y)] \hat{\pazocal{T}}_{ \kappa \lambda }[ \c4{\hat{\phi}^\dagger}(x) \c2{\hat{\phi}}(x)] \vert \c5{N,0}\rangle_{1 k} \vert \c2{N,0} \rangle_{2 l} 
    } \nonumber 
\end{gather}
\end{widetext}
Let us do the analysis using only the $\hat{\phi}^\dagger(x) \hat{\phi}(x)$ terms of the expression and not consider explicitly the terms that include $\hat{\pi}^\dagger(x) \hat{\pi}(x)$, which contribute time derivatives in the expressions. Therefore, we consider
\begin{widetext}
 \begin{gather}
    \frac{m^8}{8}\int_I d^4 x \int_I d^4 y \int_I d^4 z \int_I d^4 w \Phi(w)  \Phi(z) \Phi(y) \Phi(x) \times \\ \times \wick{
    \langle \c1{N,0} \vert_{1 i} \langle \c2{N,0} \vert_{2 j}  \c2{\hat{ \phi}^\dagger}(w) \c3{\hat{\phi}}(w)  \c1{\hat{\phi}^\dagger}(z) \c4{\hat{\phi}}(z) \c3{\hat{\phi}^\dagger}(y) \c5{\hat{\phi}}(y) \c4{\hat{\phi}^\dagger}(x) \c2{\hat{\phi}}(x) \vert \c5{N,0}\rangle_{1 k} \vert \c2{N,0} \rangle_{2 l} 
    } \nonumber 
\end{gather}
Substituting into the expression the contractions Gundhi et al. obtain  
\begin{gather}
    \frac{m^6 N^2}{32}\int_I d^4 x \int_I d^4 y \int_I d^4 z \int_I d^4 w \Phi(w)  \Phi(z) \Phi(y) \Phi(x) e^{im(z_0-x_0)} e^{im(w_0-y_0)} \tilde{\phi}_{2 l}(\vec{x}) \tilde{\phi}_{1 k}(\vec{y}) \tilde{\phi}_{1 i}(\vec{z}) \tilde{\phi}_{2 j}(\vec{w})\times \nonumber\\ \times \int \frac{d^4 s}{(2 \pi)^4} \frac{i}{s^2+m^2+i\epsilon} e^{i s (x-z)} \int \frac{d^4 r}{(2 \pi)^4} \frac{i}{r^2+m^2+i\varepsilon} e^{i r (w-y)} \label{eq:26}
\end{gather}
where the contraction $\wick{\c2{\hat{\phi}}(z) \c2{\hat{\phi}^\dagger}(x)}$ is the matter propagator $\int \frac{d^4 r}{(2 \pi)^4} \frac{i}{r^2+m^2+i\epsilon} e^{i r (x-z)}$. It is crucial to notice that the expression in Equation \ref{eq:26} consists of the product of two independent sets of triple integrals. We can rewrite the expression as 
 \begin{gather}
    \frac{m^6 N^2}{32}\left(\int_I d^4 x  \int_I d^4 z  \Phi(z)  \Phi(x) e^{im(z_0-x_0)}  \tilde{\phi}_{2 l}(\vec{x})  \tilde{\phi}_{1 i}(\vec{z})  \int \frac{d^4 k}{(2 \pi)^4} \frac{i}{k^2+m^2+i\epsilon} e^{i k (x-z)} \right) \times \nonumber\\ \times \left(\int_I d^4 y \int_I d^4 w \Phi(w) \Phi(y) e^{im(w_0-y_0)} \tilde{\phi}_{1 k}(\vec{y}) \tilde{\phi}_{2 j}(\vec{w}) \int \frac{d^4 l}{(2 \pi)^4} \frac{i}{l^2+m^2+i\varepsilon} e^{i l (w-y)}\right) 
\end{gather}
\end{widetext}
This observation is what leads Gundhi et al. \cite{gundhiCanClassicalTheories2026} to conclude that the terms $\beta^{(4)}_{ij}$ are separable. 

Nevertheless, this analysis is not correct. The substitution of the Wick contractions in Equation \ref{eq:26} is missing a crucial term. The contractions of the fields with $\ket{N,0}_{\kappa i}$ are given by Equation \ref{eq:contractket}, and it has a remaining $\ket{N-1,0}_{\kappa i}$ that has not been considered in the expression. If it is taken into account, then the expression has the extra term of $\bra{N-1,0}_{1i} \bra{N-1,0}_{2 j} \ket{N-1,0}_{1 k} \ket{N-1,0}_{2 l} $ which, for $N> 1$ evaluates to $\delta_{i k} \delta_{j l}$, exactly suppressing the off diagonal terms that would make the phase separable. The only surviving terms are in fact $\beta^{(4)}_{ij;ij}= \beta^{(4)}_{ij}$. Thus, we conclude that the analysis of Gundhi et al. is incorrect and that AH are not missing any fourth-order term in their postselection criterion. 

Nevertheless, we want to point out that even if the fourth-order terms were factorisable, the overall final state is not separable between interferometer 1 and interferometer 2. The final state would be of the form 
\begin{gather}
   \ket{\psi}=\frac{1}{2}\sum_{i} \left(1+\varphi_{1i} \right)\ket{N,0}_{1i} \otimes \sum_j \left(1+\varphi_{2j} \right) \ket{N,0}_{2j} +\nonumber \\ + \sum_{ijkl} \theta_{i l} \eta_{j k} \ket{N,0}_{1j}\otimes \ket{N,0}_{2j} 
\end{gather}
The state is not separable because there are no second-order terms $\sum_l \theta_{i l}$ and $\sum_k \eta_{j k}$ present on their own. Thus, the full state cannot be factorised in terms of interferometers 1 and 2. The reason why these second-order terms do not appear is that they would correspond to a single particle from interferometer $1$ going to interferometer $2$. Then interferometer $2$ would have $N+1$ particles and interferometer $1$ would have $N-1$. Since we are post-selecting on states that preserve $N$ particles on each interferometer, such a contribution is zero in the final state expansion. 

Therefore, even if the fourth-order term were separable, the final state is not a particle-preserving product state between the two interferometers due to the particle-preserving postselection imposed by AH.

\section{Particle entanglement or mode entanglement}
\label{sec:partmode}

We have seen that, due to postselection, the final state will not be separable between the two interferometers. Nevertheless, some replies to AH, \cite{gundhiCanClassicalTheories2026,diosiNoClassicalGravity2026} point out that the final state will be separable and no entanglement can be claimed. Concretely, these authors refer to the fact that since the interaction Hamiltonian is quadratic on quantum operators, then using the Heisenberg picture, it can be seen in analogy to \cite{diosiNoClassicalGravity2026} that defining
\begin{gather}
    \hat{a}_{\kappa i} \equiv \int d \vec{x}  \tilde{\phi}_{\kappa i}(\vec{x}) \int d \vec{p} e^{-i \vec{p} \cdot \vec{x}} \hat{a}_{\vec{p}} 
\end{gather}
then the action of the interaction unitary $\hat{U}_I(t)=\hat{\pazocal{T}} e^{-i \int_0^t d\tau \hat{H}_I(\tau)}$ will induce a particle preserving Bogoliubov transformation
\begin{gather}
    \hat{a}_{\kappa i}(t)=  \hat{U}_I(t) \hat{a}_{\kappa i} \hat{U}_I(t)^\dagger
\end{gather}
Since the initial state is separable:
\begin{equation}
    \begin{aligned}
         \ket{\psi_0}=&\frac{1}{2} \left(\left(\hat{a}_{1 L}^\dagger\right)^N + \left(\hat{a}_{1 R}^\dagger\right)^N \right)\cdot  \\ &\cdot \left(\left(\hat{a}_{2 L}^\dagger\right)^N +\left(\hat{a}_{2 R}^\dagger\right)^N \right) \ket{0}
    \end{aligned}
\end{equation}
The final state is also:
\begin{equation}
    \begin{aligned}
        \ket{\psi(t)}=& \frac{1}{2} \left(\left(\hat{a}_{1 L}^\dagger(t)\right)^N + \left(\hat{a}_{1 R}^\dagger(t)\right)^N \right)\cdot\\ &\cdot \left(\left(\hat{a}_{2 L}^\dagger(t)\right)^N +\left(\hat{a}_{2 R}^\dagger(t)\right)^N \right) \ket{0}
    \end{aligned}
\end{equation}

How can these two claims be reconciled? The answer is that they correspond to two distinct types of separability, and thus two different types of entanglement, present in indistinguishable particle systems. One is the separability between the two interferometers, which corresponds to the notion of mode entanglement \cite{zanardiQuantumEntanglementFermionic2002,friisFermionicmodeEntanglementQuantum2013,shapourianEntanglementNegativityFermions2019,benattiEntanglementFermionSystems2014,shiQuantumEntanglementIdentical2003,tibauvidalQuantumOperationsInformation2021}. The other refers to the separability of the $2N$ particles, in the context of particle entanglement \cite{schliemannQuantumCorrelationsTwofermion2001,eckertQuantumCorrelationsSystems2002,wisemanEntanglementIndistinguishableParticles2003,ghirardiGeneralCriterionEntanglement2004,plastinoSeparabilityCriteriaEntanglement2009,ieminiQuantumnessCorrelationsIndistinguishable2014,debarbaQuantumnessCorrelationsFermionic2017,sarosiEntanglementClassificationThree2014}. 

The first notion is the one measured in the crude BMV protocol. The entanglement we want to observe is between the two interferometers located in two different spatial regions. The particle notion refers to the individual particles in the experiment and whether they have become entangled with each other. In AH's calculation, since particles are allowed to leak from one interferometer to the other, the final state is $ 2N$-particle separable but still mode-entangled. If the $N$ particles in each interferometer are required to remain in their respective interferometers, then it follows that if the final state is $2N$-particle separable, it is also interferometer-mode separable. However, since we have this contamination leakage effect, entanglement between the two interferometers can be created and measured. 

We believe that any implementation of the BMV protocol with indistinguishable particles should be shielded from this effect and should not allow leakage from one interferometer to the other, thereby ensuring that semiclassical gravity cannot generate entanglement. 

To fully exemplify these differences, we consider the simple case where $N=1$ and consider some simple dynamics such that 
\begin{align}
    &\hat{a}_{1 L}(t_f)=\frac{1}{\sqrt{2}} \left(\hat{a}_{1 L}+ \hat{a}_{2 L} \right) \\ &\hat{a}_{2 L}(t_f)=\frac{1}{\sqrt{2}} \left(\hat{a}_{1 L}- \hat{a}_{2 L} \right) \\ &\hat{a}_{1 R}(t_f)=\hat{a}_{1 R} \\ &\hat{a}_{2 R}(t_f)= \hat{a}_{2 R} 
\end{align}
Now we can see that the final state has the product form
\begin{equation}
    \begin{aligned}
        \ket{\psi(t_f)}=&\frac{1}{4} \left(\hat{a}_{1 L}^\dagger+\hat{a}_{2 L}^\dagger+\sqrt{2} \hat{a}_{1 R}^\dagger\right) \cdot\\ &\cdot \left(\hat{a}_{1 L}^\dagger-\hat{a}_{2 L}^\dagger+\sqrt{2}\hat{a}_{2 R}^\dagger\right) \ket{0}
    \end{aligned}
\end{equation}
but in terms of the separation between the interferometers, we do not have a product structure
\begin{equation}
    \begin{aligned}
        \ket{\psi(t_f)}&=\frac{1}{4} \left[\left(\hat{a}_{1 L}^\dagger+\sqrt{2} \hat{a}_{1 R}^\dagger\right)\hat{a}_{1 L}^\dagger+\right.\\ &+\hat{a}_{2 L}^\dagger\left(-\hat{a}_{2 L}^\dagger+\sqrt{2} \hat{a}_{2 R}^\dagger\right) +\\&\left.+\sqrt{2}\left(\hat{a}_{1 L}^\dagger\hat{a}^\dagger_{2R}-\hat{a}_{1 R}^\dagger\hat{a}^\dagger_{2 L}+\sqrt{2} \hat{a}_{1 R}^\dagger\hat{a}^\dagger_{2 R} \right)\right] \ket{0}
    \end{aligned}
\end{equation}
Notice that in the last expression the first two terms correspond to both particles being in the first interferometer and the second, respectively. If we apply the postselection criterion here, then our final state would be 
\begin{gather}
    \ket{\psi(t_f)} \approx \frac{1}{2}\left(\hat{a}_{1 L}^\dagger\hat{a}^\dagger_{2R}-\hat{a}_{1 R}^\dagger\hat{a}^\dagger_{2 L}+\sqrt{2} \hat{a}_{1 R}^\dagger\hat{a}^\dagger_{2 R} \right) \ket{0}
\end{gather}
which is an entangled state between the two interferometers. 

This simple example shows how the entanglement between the two interferometers can be present despite having the product structure at the particle level given by the quadratic Hamiltonian. Nevertheless, it shows that it is a direct result of particles from one interferometer having a nonzero probability of travelling to a branch of the other interferometer. 

\section{Bringing spin back} \label{sec:spin}
The identified mechanism of exchanging leaked particles between the two interferometers, contaminating one another, would pose problems for the final analysis if spin is considered. The spin degree of freedom could partially distinguish the particles from the two interferometers; thus, it would undermine part of the claimed entangling effect.  

AH removes spin from their analysis for simplicity. They assume that the generated phases in the spatial degrees of freedom will be transferred to the spin degrees of freedom of the interferometer branches.

However, in our view, it is unreasonable to assume that the spin degree of freedom will remain unaffected when particles are physically leaking from the interferometer branches. In these cases, spin is either a collective mode of the $N$ particles that depends on their spatial properties such as their distributions and wavefunction shape, or is the sum of individual spin degrees of freedom attached to each particle. 

If it is the first case, one would need to prove that if one particle leaks out so far away to another interferometer branch that the collective spin mode would remain completely unchanged. We focus instead on the second case for simplicity, where one of the $N$ spins leaves the interferometer branch and is replaced by another spin from the other interferometer. 

Let us consider that left branches are prepared with all $N$ spins up $\ket{\uparrow_N}$, and right branches with all $N$ spins down, $\ket{ \downarrow_N}$ . Then, our initial state adding spin would be 
\begin{equation}
    \begin{aligned}
        \ket{\eta_0}&=\frac{1}{2}\Big( \ket{N,0}_{1 L}\ket{\uparrow_N} \ket{0,N}_{2 L} \ket{\uparrow_N} +\\&+\ket{N,0}_{1 R} \ket{ \downarrow_N}\ket{N,0}_{2 L} \ket{\uparrow_N}+ \\ &+\ket{N,0}_{1 L} \ket{\uparrow_N}\ket{0,N}_{2 R} \ket{ \downarrow_N}+\\&+\ket{N,0}_{1 R} \ket{ \downarrow_N} \ket{N,0}_{ 2 R} \ket{ \downarrow_N}\Big) 
    \end{aligned}
\end{equation}

However, if we focus on the $\beta^{(4)}_{ij}$ terms that give rise to the claimed entanglement in the postselected final state, we obtain certain final-state components in which the $N$ spins are not completely aligned. Concretely, these would be the terms $\beta_{L R}^{(4)},\beta_{R L}^{(4)}$ that would have the respective transition $\ket{N,0}_{1 L} \ket{\uparrow_N}\ket{0,N}_{2 R} \ket{ \downarrow_N}$ to the term given by $\ket{N,0}_{1 L} \ket{\uparrow_{N-1} \downarrow}\ket{N,0}_{2 R} \ket{ \downarrow_{N-1} \uparrow}$ and the transition from $\ket{N,0}_{1 R} \ket{ \downarrow_N}\ket{0,N}_{2 L} \ket{\uparrow_N}$ to $\ket{N,0}_{1 R} \ket{ \downarrow_{N-1} \uparrow}\ket{0,N}_{2 L} \ket{\uparrow_{N-1} \downarrow}$.

These terms are a direct consequence of the entangling mechanism consisting of leaking particles from one interferometer to the other. The fact that such leaking is in the spatial degrees of freedom means that their individual spin degrees of freedom remain unaltered during the process. 

If we apply the same logic as for the non-particle-preserving states, such components should be postselected away, since the spin values we are encoding our logical qubit in are $\ket{\uparrow_N}$ and $\ket{\downarrow_N}$. We can rephrase the result by noting that, in these components, spin distinguishes the particles in both interferometers, which is why their contributions vanish. Therefore, the only surviving entangling contributions would be $\beta^{(4)}_{L L }, \beta^{(4)}_{R R }$, which, due to the geometry of the system, have the same value. 

This point proves that the effective entangling terms would be smaller than in the AH analysis. The terms are a consequence of having indistinguishable particles and not shielding the interferometers correctly, allowing for contamination due to leakage from one another. 

\section{Discussion}\label{sec:discussion}
In this article, we explore AH's claims that classical gravity can produce entanglement in quantum systems. We have analysed the interpretation of their results using the same formalism of quantum field theory in curved spacetime. 

The main conclusion we extract from our comments is that classical gravity does not mediate the entanglement. We agree that in the AH setup, the field excitations within the two interferometers will become entangled in the postselected state. The process by which they become entangled arises from the leakage of excitations from one interferometer to the other and the indistinguishable nature of the starting excitations within the two interferometers. 

We conclude that under classical gravity, two indistinguishable $N$-particle states would be able to become entangled if there is no barrier between the two interferometers preventing the diffusion of particles from one interferometer to the other and a protocol of postselection to reasonable states is enforced. 

We have shown how, if the states are distinguishable, the entangling contribution calculated in AH vanishes. Moreover, we have seen that when taking spin into account, due to partial distinguishability, part of the entangling contributions also vanish. Therefore, it is clear that the entangling contributions in AH arise due to the indistinguishable nature of the particles involved in the two interferometers. 

Moreover, we have seen how the exact entangling contributions from AH arise from the exchange of two real particles within the two interferometers. The authors in AH claim that the entanglement arises due to a virtual mediation given by the quantum matter field. We believe this interpretation is misleading, since entanglement is not mediated in this scenario. Instead, due to the gravitational classical non-flat potential, the wave packets associated with the particles within the interferometers can diffuse freely towards the other interferometer, since no barrier is modelled. Then, due to the indistinguishable nature of the particles and the non-separability of the probability amplitudes of a mutual leakage from one interferometer arm to the other, such contamination arises as an entangling contribution. There is no mediation, just direct contact and confabulation of the two systems within the interferometers due to their indistinguishable nature. Furthermore, we have seen that the entangling terms are not given as a result of a virtual process. The contribution to the entangling term is given by collapsing all the virtual contributions to zero and just picking up the on-shell term that refers to a real particle travelling from one interferometer arm to another. 

Such an analysis leads us to believe that the AH effect should be completely mitigated in any desirable experimental implementation of the BMV protocol. In any experimental design for such a protocol, measures would already be in place to prevent cross-contamination of particles from one interferometer leaking into the other, such as trapping potentials or barriers \cite{dipietraBoseMarlettoVedralExperimentNanodiamond2024,aspelmeyerWhenZehMeets2022}. Another method to prevent this undesirable effect for BMV would be to use distinguishable particles. We have proven using AH's own formalism that if the particles of interferometer 1 are distinguishable from interferometer 2, then classical gravity cannot produce entanglement between them at all. 

In the article, we also refer to several replies to the AH claim, in which we point out two misconceptions. The first is that the entangling phase is not separable, and thus the final state has no entanglement between the two interferometers. We have shown that the analysis in \cite{gundhiCanClassicalTheories2026} was missing the contribution of the bras and kets in the Wick contractions with the state, which led to the off-diagonal terms being zero and thus not making the entangling contribution separable. Moreover, we have seen that even if it were separable, the final postselected state would not be, and some entanglement would still be present. 

The second is that the claim that the final state is separable when solved exactly, given in \cite{diosiNoClassicalGravity2026}, is true but refers to another type of entanglement. They refer to particle entanglement and not to entanglement between the two interferometers, which is the relevant one in this experimental setting. The entanglement between the two interferometers is the one being measured and claimed to be a signature of quantum gravity in the BMV protocols. We have shown how a state can be particle-separable yet entangled in the postselected state between the two interferometers. Again, such a distinction is only relevant when particles are indistinguishable, and one cannot track their identity within the interferometers. Therefore, despite the state being particle separable, entanglement between the two interferometers can be present in the postselected particle-preserving state.  

Our article sheds light on the AH effect and its entangling term by precisely describing the physical processes that underlie them. We are then well-positioned to comment on whether the theoretical claims in BMV conflict with this result. We believe there is no theoretical clash. The BMV theoretical underpinning always assumes that the two quantum systems are distinguishable and free of contamination. Their concepts rely on the notion of gravitational mediation of entanglement, which the AH effect does not display. Since entanglement is generated, it is natural to ask which component of LOCC (Local Operations and Classical Communication) is broken. We believe both LO (Local Operations) and CC (Classical communication) are broken. In an analogy of the AH effect to a finite particle system of Alice and Bob each holding $N$ qubits, the AH entangling term consists of having a small probability of Alice and Bob both coordinating to exchange one of their qubits. Such a protocol violates LO since it is a global coordination between Alice and Bob and does not happen independently of each other. Going back to the AH effect, this global nature arises when imposing the postselection restriction of discarding states in the interferometers that have lost particles. The protocol also violates CC because Alice and Bob are exchanging physical qubits, which are not classical.

\section*{Aknowledgements}
We want to thank Ali Akil, Vlatko Vedral, and Chiara Marletto for fruitful discussions on this topic and related ideas. A.I. and N.T.V. acknowledge the support of UK Research and Innovation (UKRI) through the Future Leaders Fellowship.  

\bibliographystyle{unsrt} 
\bibliography{zoteroN}

\end{document}